\documentclass[conference]{IEEEtran}
\IEEEoverridecommandlockouts
\usepackage{lipsum}
\usepackage{mathtools}
\usepackage{cuted}
\usepackage{mwe}
\usepackage[caption=false, font=footnotesize]{subfig}
\usepackage{mathtools}

\usepackage{graphicx}
\usepackage{microtype}
\usepackage{balance}
\usepackage{cite}
\usepackage{amsmath,amssymb,amsfonts}
\usepackage{algorithmic}
\usepackage{textcomp}
\usepackage{xcolor}
\usepackage{vmr-symbols-vecbold}
\usepackage{standard-macros}

\usepackage[pdfborder={0 0 0}]{hyperref}% For email addresses

\DeclarePairedDelimiter\norm{\lVert}{\rVert}%
\makeatletter
\def\@IEEEinterspaceratioM{0.265}
\def\@IEEEinterspaceMINratioM{0.1651}
\def\@IEEEinterspaceMAXratioM{0.38}

\def\@IEEEinterspaceratioB{0.31}
\def\@IEEEinterspaceMINratioB{0.19}
\def\@IEEEinterspaceMAXratioB{0.38}
\@IEEEtunefonts
\makeatother
\def\BibTeX{{\rm B\kern-.05em{\sc i\kern-.025em b}\kern-.08em
    T\kern-.1667em\lower.7ex\hbox{E}\kern-.125emX}}

\begin{document}

\author{
\IEEEauthorblockN{Yasaman Ettefagh$^\text{1}$, Sven Jacobsson$^\text{1,2}$, Anzhong Hu$^\text{3}$, Giuseppe Durisi$^\text{1}$, and Christoph Studer$^\text{4}$}     
\IEEEauthorblockA{$^\text{1}$Department of Electrical Engineering, Chalmers University of Technology, Gothenburg, Sweden}
\IEEEauthorblockA{$^\text{2}$Ericsson Research, Gothenburg, Sweden}
\IEEEauthorblockA{$^\text{3}$School of Communication Engineering, Hangzhou Dianzi University, Hangzhou, China}
\IEEEauthorblockA{$^\text{4}$Cornell Tech, New York, NY}
% Chalmers University of Technology, Gothenburg, Sweden} 
% \IEEEauthorblockA{$^\text{2}$Ericsson Research, Gothenburg, Sweden} 
% \IEEEauthorblockA{$^\text{3}${School of Communication Engineering,
% Hangzhou Dianzi University, Hangzhou, China} 
% \IEEEauthorblockA{$^\text{4}$Cornell Tech, New York, NY}
% %\vspace{-0.1 in}
 \thanks{The work of YE, SJ, and GD was supported in part by the Swedish Foundation for Strategic Research under grant ID14-0022, and by the Swedish Governmental Agency for Innovation Systems (VINNOVA) within the competence center ChaseOn. The work of CS was supported by Xilinx Inc.\ and by the US NSF under grants ECCS-1408006, CCF-1535897,  CCF-1652065, CNS-1717559, and ECCS-1824379.}
}

\title{All-Digital Massive MIMO Uplink and Downlink Rates under a Fronthaul Constraint}
\maketitle
\begin{abstract}
We characterize the rate achievable in a bidirectional  quasi-static link where several user equipments communicate with a massive multiple-input multiple-output base station (BS).
In the considered setup, the BS operates in full-digital mode, the physical size  of the antenna array is limited, and there exists a rate constraint on the fronthaul interface connecting the (possibly remote) radio head to the digital baseband processing unit.
Our analysis enables us to determine the optimal resolution of the analog-to-digital and digital-to-analog converters as well as the optimal number of active antenna elements to be used in order to maximize the transmission rate on the bidirectional link,  for a given constraint on the outage probability and on the fronthaul rate. 
We investigate both the case in which perfect channel-state information is available, and the case in which channel-state information is acquired through pilot transmission, and is, hence, imperfect.
For the second case, we present a novel rate expression that relies on the generalized mutual-information framework.
\end{abstract}

%\begin{IEEEkeywords}
%Massive MIMO, fronthaul, low-resolution quantizers, CSI acquisition.
%\end{IEEEkeywords}

\section{Introduction}
Full-digital massive multiple-input multiple-output (MIMO) architectures involve transferring a significant amount of data from  the base-band unit (BBU) to the the remote radio head (RRH), which consists of antennas and radio-frequency (RF) circuitry.
This is particularly challenging for massive MIMO systems operating in the millimeter wave part of the spectrum because of the larger chunks of bandwidth available at such frequencies. 
One additional challenge is that the RRH and the BBU are often at different locations, for accessibility, maintenance, and reconfigurability purposes.
Hence, they need to be connected through a fronthaul interface whose capacity is finite and typically a limiting design factor.
Innovative deployments may even involve distributed architectures in which not even the antenna elements are co-located~\cite{sezgin19-07a}.

In order to understand the scale of this interconnect data rate problem, consider for example a base station (BS) with $500$ active antenna elements, each one connected to two $10$-bit data converters (one for the real and one for the imaginary part of the complex baseband signal) that operate at $1$\,GS/s. 
Such an architecture produces $10$\,Tb/s of raw baseband data, which is in excess of what current fronthaul links can support. 

The problem of fronthaul data compression is well-investigated in the cloud radio access network literature (see, e.g.,~\cite{park14-11a} and references therein).   
However, the focus of that line of work is on advanced solutions that rely on multiterminal compression techniques and require significant signal-processing capabilities at the RRH. In contrast, the focus of this paper is on light-weight solutions that are relevant for low-complexity and, hence, low-cost RRHs.
Specifically, we consider the simple approach of reducing the required fronthaul data rate by lowering the precision of the analog-to-digital (ADC) and digital-to-analog (DAC) converters at the RRH. 

The theoretical performance achievable with massive MIMO architectures for the case in which the BS antennas are co-located and each RF chain is connected to low-resolution ADCs/DACs has been the subject of many recent investigations in the scientific literature.
These include the characterization of the information-theoretic achievable rates in the ergodic scenario~\cite{jacobsson17-06a,li17-08a,mollen17-01a,bjornson19-02b}, the design of channel-estimation and data-detection algorithms~\cite{li17-08a,studer16-06a}, and of both linear and nonlinear precoders~\cite{jacobsson17-11a,jacobsson19-03b}. 
All these results indicate that satisfactory performance can be achieved even when equipping the BS with $1$-bit data converters.
Furthermore, using data converters with $3$--$5$ bits of resolution is typically sufficient to approach infinite-precision performance.

%\todo{Continue with short literature review about massive MIMO with low-prec quantizers}

\paragraph*{Contributions}
We consider multi-user bidirectional massive MIMO communication over a quasi-static channel.
For a given number of BS antennas, and a given resolution of the data converters, we determine a lower bound on the rate that can be supported in both the uplink and the downlink for a given constraint on the outage probability, i.e., the probability that either the uplink or the downlink is in outage. 
We then use this bound to investigate whether, for a given fronthaul constraint, it is better to use a large number of  antennas connected with low-precision data converters, or a smaller number of antennas connected with high-precision data converters. 

Using the Bussgang-based approach described in~\cite{jacobsson18-06a}, we first investigate the case in which perfect channel-state information (CSI) is available at the BS and at the user equipments (UEs).
This assumption is standard in analyses over quasi-static channels for the case of infinite-precision quantizers. 
Indeed, CSI can be acquired with arbitrary accuracy in the outage regime without any rate penalty~\cite[p. 2632]{biglieri98-10a}.  
However, as we will show in this paper, the validity of this assumption is questionable when the BS is equipped with low-precision ADCs. 
Motivated by this observation, we extend our analysis to  the case in which the available CSI at the BS is noisy.
Since, under imperfect CSI, the rate bound presented in~\cite{jacobsson18-06a} does not hold, we derive a novel bound, based on~\cite{lapidoth02-05a}, which relies on scaled nearest-neighbor decoding and on the generalized mutual-information framework. 

Numerical results, for the case in which the BS antennas are modeled as a fixed-size uniform linear array (ULA) and the channel is line-of-sight (LOS), reveal that architectures involving a large number of antennas and low-precision data converters (with as few as $2$-bit resolution) are preferable to solutions involving a small number of antennas that rely on  higher-precision data converters even when limited CSI accuracy is accounted for.

\section{System Model} \label{sysmodel}
\begin{figure}[tp]
\centering
\subfloat[Uplink]{
% : the RRH quantizes the received signal using $2B$ $Q$-bit ADCs; the quantized I/Q data are sent to the BBU over the bandwidth-constrained fronthaul link.
\includegraphics[width=.7\columnwidth]{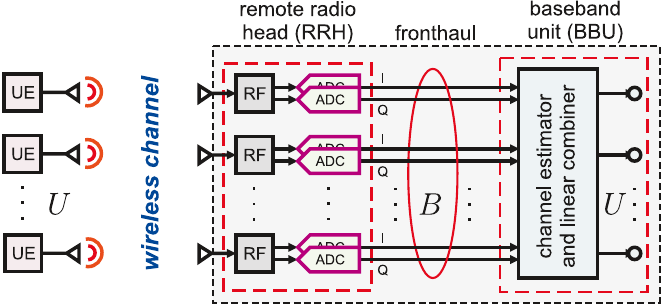}\label{fig:system_uplink}} \\
\subfloat[Downlink]{
% :  the BBU quantizes the  precoded vector to match the precision of the $2B$ $Q$-bit DACs at the RRH; the I/Q data are sent to the RRH over the bandwidth-constrained fronthaul~link.
\includegraphics[width=.7\columnwidth]{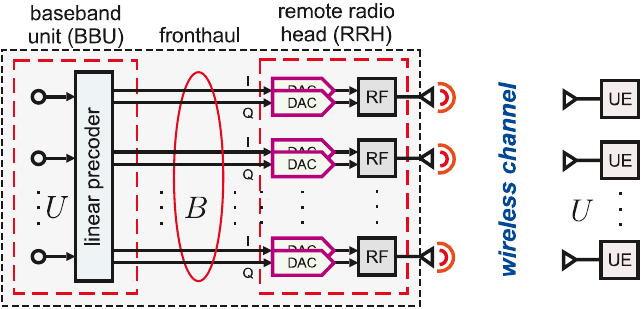}\label{fig:system_downlink}}
\caption{Overview of the low-resolution all-digital $B$-antenna BS architecture considered in the paper. 
The BS consists of a BBU and a RRH that are connected via a rate-constrained fronthaul interface.}
\label{fig:system}
\end{figure}

We consider a multi-user scenario where a BS equipped with $B$ antennas communicates with $U\ll B$ UEs.
As depicted in Fig.~\ref{fig:system}, the BS consists of a RRH and a BBU that are connected via a rate-constrained fronthaul interface. 
Each antenna is equipped with a pair of $Q$-bit converters, one for the in-phase and one for the quadrature component. 
In the uplink, the signal transmitted by the $U$ UEs is quantized at the $B$ BS antennas using the low-precision data converters.
The quantized signal is then transferred to the BBU via the fronthaul link, where channel estimation, linear combining, and decoding are performed.
In the downlink, the linearly-precoded signal is quantized at the BBU and transferred over the fronthaul link, where it is converted into the analog domain and transmitted over the $B$ antennas.  
It follows that an architecture with $Q$-bit converters and $B$ active antennas requires a fronthaul interface able to operate at $2BQ$\,bit/s/Hz.

We assume uniform, symmetric, mid-rise quantizers with step size $\Delta$ and $Q$-bit resolution.
Specifically, let $r\in\reals$ be the input of the quantizer.
Then, the output $\setQ(r)$ is given by
\begin{align} \label{eq:quantizer_uniform}
    \mathcal{Q}(r) &= 
    \begin{cases}
    \frac{\Delta}{2}(1 - L) & \text{if } r < - \frac{\Delta}{2}L \\
    \Delta \lfloor \frac{r}{\Delta}\rfloor + \frac{\Delta}{2} & \text{if } - \frac{\Delta}{2}L  \le r < \frac{\Delta}{2}L \\
    \frac{\Delta}{2}(L - 1) & \text{if } r \ge \frac{\Delta}{2}L.
    \end{cases}	
\end{align}
Here, $L=2^Q$ denotes the number of quantization levels.
For a complex-valued input $z$, we let $\setQ(z)=\mathcal{Q}(\Re\{z\}) + j \mathcal{Q}(\Im\{z\})$. For a vector $\vecz$, we denote by $\setQ(\vecz)$ the result of applying $\setQ(\cdot)$ entrywise to its elements.

We now provide an expression for the rate achievable in the uplink and downlink for a given channel realization, which is assumed to stay constant over the duration of each codeword (i.e., we consider the outage regime). 
We first focus on the case of perfect CSI at the BS, and then move to the case of imperfect~CSI.

\section{Achievable Rate: Perfect CSI}\label{sec:ar_perfect_csi}
As discussed in~\cite{jacobsson18-06a}, a lower bound on the rates achievable with Gaussian codebooks, both in the uplink and in the downlink, can be obtained using Bussgang's theorem~\cite{bussgang52a}. 
This result will be reviewed next as it will be useful for our analysis of the imperfect-CSI case in Section~\ref{sec:ar-imperfect-csi}.

\paragraph*{Uplink}
The $B$-dimensional received signal at the BS  for the $k$th discrete-time channel use can be written as 
\begin{align}
    \mathbf{y}^\text{ul}_k 
    &= \mathbf{H} \textbf{s}^\text{ul}_k + \mathbf{n}^\text{ul}_k. 
\end{align}
Here, $\textbf{s}^\text{ul}_k = [s^\text{ul}_{k,1},\dots, s^\text{ul}_{k,U}]^T\in \mathbb{C}^U$ is the transmitted signal from the $U$ single-antenna UEs.
We assume that the $\{s^\text{ul}_{k,u}\}$ are drawn independently from a $\jpg(0,\snr\sub{ul})$ distribution, where  $\snr\sub{ul}$ refers to the uplink power.
The vector $\mathbf{n}^\text{ul}_k\sim \mathcal{CN}(\textbf{0}_B,\mathbf{I}_B)$ is the AWGN at the BS. 
Finally, $\matH\in \complexset^{B\times U}$ denotes the channel matrix.
Note that the matrix stays constant across channel uses---indeed, it does not depend on $k$ (outage setup).
In this section, we assume that $\matH$ is known to the BS; in Section~\ref{sec:acquring_csi}, we will discuss the validity of this hypothesis.
We model the output of the ADCs as follows:\footnote{In the remainder of this section, we drop the index $k$ to keep the notation compact.}
\begin{align}
    \mathbf{r}^\text{ul} 
    &= \mathcal{Q}\lefto(\mathbf{A}\mathbf{y}^\text{ul}\right). 
\end{align} 
Here, $\matA$ is a diagonal matrix that models the effect of the automatic-gain-controller circuit, whose function is to scale the input of the quantizer so that it matches its dynamic range. 
This scaling matrix is given by 
\begin{equation}
    \mathbf{A} = \frac{1} {\sqrt{B}}\mathrm{diag}(\mathbf{C}_{\textbf{y}^\text{ul}})^{-1/2}
\end{equation}
where $\matC_{\textbf{y}^\text{ul}}=\rho_\text{ul}\mathbf{H}\mathbf{H}^H+\mathbf{I}_B$ denotes the covariance matrix of the quantizer input.
With this choice of $\matA$, we ensure that the average power of the received signal at each antenna is $1/B$. 

A key step to obtain a lower bound on the achievable rates is to write the quantized channel output as the sum of the linear minimum mean-square error (LMMSE) estimate of $\mathbf{r}^\text{ul}$ given the quantizer input $\mathbf{A}\mathbf{y}^\text{ul}$ plus the uncorrelated estimation error~$\mathbf{e^\text{ul}}$ as follows:
\begin{equation}
    \mathbf{r}^\text{ul}=\mathbf{G^\text{ul}Ay^\text{ul}}+\mathbf{e^\text{ul}}.
\end{equation}
Here, $\matG^\text{ul}$ is the LMMSE filter matrix. 
Since the quantizer input is conditionally Gaussian given $\matH$, this filter takes a particularly simple form. 
Specifically, it follows from Bussgang's theorem~\cite{bussgang52a} that $\matG^\text{ul}$ is diagonal and given by~\cite{jacobsson18-06a}:
\begin{align}
    \mathbf{G^\text{ul}}=& \frac{\Delta}{\sqrt{\pi}}  \operatorname{diag}(\mathbf{A}\mathbf{C}_{\textbf{y}^\text{ul}}\mathbf{A})^{-1/2} \nonumber \\
    & \times \sum_{i=1}^{L-1} \exp\left(-\Delta^2 (i-L/2)^2 \operatorname{diag}(\mathbf{A}\mathbf{C}_{\textbf{y}^\text{ul}}\mathbf{A})^{-1}\right).
  \end{align}

We assume that the BBU obtains an estimate $\hat \vecs^\text{ul}$ of $\vecs^\text{ul}$ by means of a linear combiner $\matW\in\complexset^{B\times U}$.
Specifically, the BBU computes $\hat\vecs^\text{ul}=\mathbf{W}^H\textbf{r}^\text{ul}$.
For the case of maximum-ratio (MR) combining, we have for example that $\matW=\matG\matA\matH$. 

Decomposing the estimate $\hat s^\text{ul}_u$ of $s^\text{ul}_u$ into the sum of useful signal, residual multiuser interference, quantization noise, and additive noise, we can write  the signal to interference noise and distortion ratio (SINDR) $\gamma_u^\text{ul}$ as follows:
\begin{IEEEeqnarray}{rCl} \label{eq:sindr_uplink}
    \gamma_u^\text{ul}  &=& \frac{ \rho_\text{ul} \abs{\mathbf{w}_u^H \mathbf{G}^\text{ul} \mathbf{A} \mathbf{h}_u}^2}{\rho_\text{ul}\sum\limits_{v \neq u} \abs{\mathbf{w}_u^H \mathbf{G}^\text{ul} \mathbf{A} \mathbf{h}_v}^2 + \mathbf{w}_u^H\mathbf{C}_{\mathbf{e}^\text{ul}}\mathbf{w}_u +\norm{\mathbf{A}\mathbf{G}^\text{ul}\mathbf{w}_u}^2}. \IEEEeqnarraynumspace
\end{IEEEeqnarray}
Here, $\matC_{\vece^\text{ul}}$ denotes the correlation matrix of the distortion error. 
This matrix can be computed in closed form using the arcsine law~\cite{van-vleck66a} for the case $Q=1$.
For $Q>1$, no closed-form expression is available but accurate, yet easy-to-compute approximations are given in~\cite{jacobsson19-03b}.
A linear combiner, dubbed distortion-aware MMSE combiner, which maximizes~\eqref{eq:sindr_uplink}, was recently proposed in~\cite{bjornson19-02b}.

It follows from \cite[App.~B]{lapidoth02-05a} that, for a given $\matH$, the uplink achievable rate for the perfect CSI case can be lower-bounded by $\log(1+\gamma_u^\text{ul})$.
Furthermore, this rate is achieved by a mismatched nearest-neighbor decoder that treats residual interference and quantization errors as Gaussian noise. 
For a target rate $R$, the uplink outage probability is given by 
\begin{equation} \label{eq:ULOUt}
    \text{Pr} \{ \operatorname{log}(1+\gamma_u^\text{ul}) < {R} \}.
\end{equation}

\paragraph*{Downlink}
Following steps similar to the uplink case, we can write the downlink SINDR~$\gamma_u^\text{dl}$ as follows:
\begin{IEEEeqnarray}{rCl} \label{eq:sindr_downlink}
\gamma_u^\text{dl}  &=& \frac{ \rho_\text{dl} \abs{ \mathbf{h}_u^T  \mathbf{G}^\text{dl} \mathbf{p}_u}^2}{\rho_\text{dl}\sum\limits_{v \neq u} \abs{\mathbf{h}_u^T \mathbf{G}^\text{dl}\mathbf{p}_v}^2 + \mathbf{h}_u^T\mathbf{C}_{\mathbf{e}^\text{dl}}\vech^*_u +1}. \IEEEeqnarraynumspace
\end{IEEEeqnarray}
Here, $\matG^\text{dl}$ is the downlink Bussgang gain, $\mathbf{C}_{\mathbf{e}^{\text{dl}}}$ is the downlink quantization error, and $\vecp_u$ is the $u$th column of the precoding matrix (see~\cite{jacobsson18-06a} for the details).
Under the assumption that $\matH$ and $\mathbf{h}_u^T  \mathbf{G}^\text{dl} \mathbf{p}_u$ are perfectly known to the BS and to UE $u$th, respectively, the achievable downlink rate for a given~$\matH$ can be lower-bounded by $\log(1+\gamma_u^\text{dl})$.
Again, this bound is achieved by a mismatched nearest-neighbor decoding rule that treats residual interference and quantization errors as Gaussian noise. 
For a target rate $R$, the downlink outage probability is given by 
\begin{equation} \label{eq:ULOUt}
    \text{Pr} \{ \operatorname{log}(1+\gamma_u^\text{dl}) < {R} \}.
\end{equation}

\section{Acquiring CSI}\label{sec:acquring_csi}
For the quasi-static scenario considered in this paper, acquiring arbitrarily precise CSI can be achieved with no rate penalty in the large-blocklength regime implicitly considered in outage probability analyses, for the case in which the ADCs have infinite precision.
Indeed, it is sufficient to let the number of pilot symbols grow sublinearly with the blocklength.
However, when the ADCs have limited precision, the estimated CSI may remain imperfect, independently on how many pilot symbols are~used.

\begin{figure}[t]
    \centering
    \includegraphics[width=.8\columnwidth]{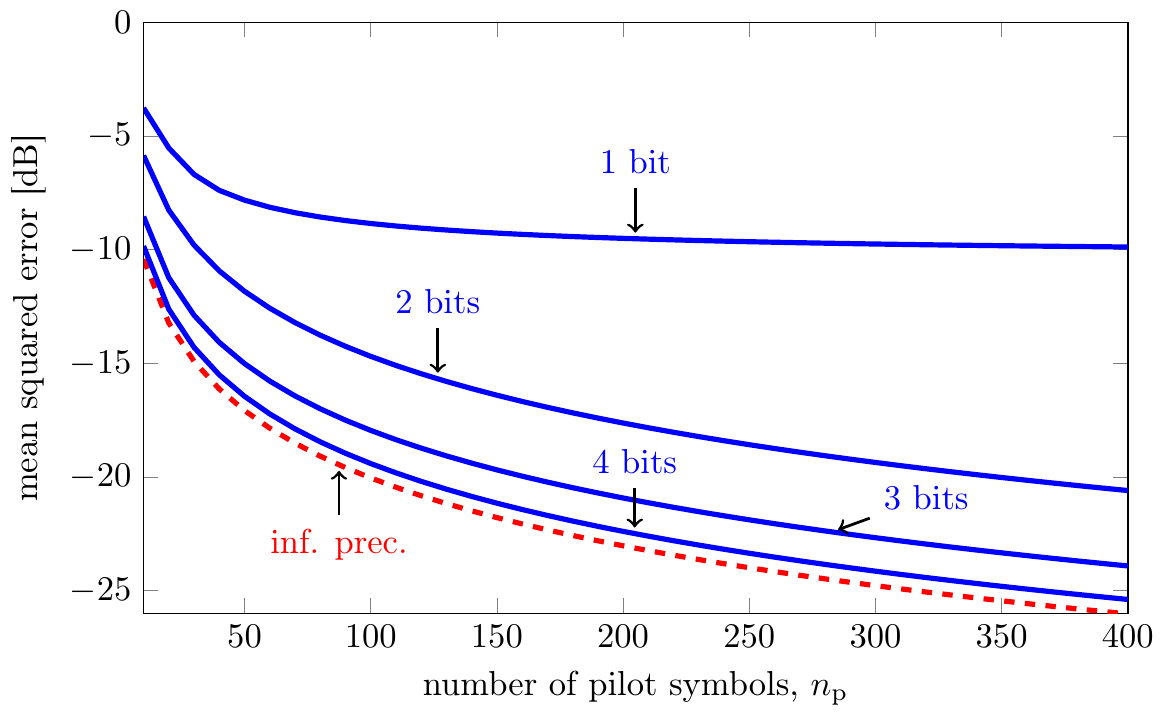}
    \caption{MSE for the Bussgang MMSE channel estimator versus number of pilot symbols; $B=100$, $U=10$, $\rho\sub{ul}=10$, i.i.d. Rayleigh-fading channel.}
    \label{fig:ChEstNp}
\end{figure}

Consider, for example, the case of an \iid Rayleigh fading channel. 
Assume that the $U$ UEs transmit $n\sub{p}$ orthogonal pilot sequences to estimate the channel. 
Bussgang's decomposition allows one to obtain a simple linear channel estimator that is based on the MMSE principle.
Such an estimator, referred to as Bussgang MMSE, was proposed in~\cite{li17-08a} for the case of 1-bit quantizers and later generalized in~\cite{jacobsson17-06a} to quantizers with arbitrary resolution. 
The mean-square error (MSE) of this Bussgang MMSE estimator as a function of $n\sub{p}$ is depicted in Fig.~\ref{fig:ChEstNp}. 
Here, $U=10$, $B=100$, $\rho\sub{ul}=10\dB$, and the pilot matrix is a submatrix of the discrete Fourier transform (DFT) matrix. 
To compute the covariance matrix of the quantization error, which is needed for the evaluation of the MSE, we used the diagonal approximation proposed in~\cite{jacobsson19-03b}. This approximation is  accurate for the parameters considered in the figure. 
As illustrated in Fig.~\ref{fig:ChEstNp}, the MSE  for the case of 1-bit quantization saturates as $n\sub{p}$ grows.
However, the gap to the infinite-precision curve decreases rapidly as the number of quantization bits increase.

To investigate the dependence of the MSE on $\rho\sub{ul}$, we now focus  on the 1-bit case and assume that $U=1$.
Under these assumptions, one can show that the MSE takes the following simple form:
\begin{multline}\label{eq:mmse_1bit}
    \mathrm{MSE}(n_\text{p},\rho_\text{ul})= \\1-\frac{\rho_\text{ul}}{1+\rho_\text{ul}} \left(\frac{n_\text{p}}{\frac{\pi}{2}+(n_\text{p}-1)\arcsin\left(\frac{\rho_\text{ul}}{1+\rho_\text{ul}}\right)}\right).
\end{multline}
It is key to observe that
\begin{equation}
    \lim_{n\sub{p}\to\infty} 
    \mathrm{MSE}(n\sub{p},\rho_\text{ul})
    = 1 - \frac{ \frac{\rho_\text{ul}}{1+\rho_\text{ul}} }{ \arcsin\!\left(\frac{\rho_\text{ul}}{1+\rho_\text{ul}}\right)}
\end{equation}
which implies that the MSE remains bounded above zero independently of the number of pilot symbols.

\begin{figure}[t]
  \centering
  \includegraphics[width=.8\columnwidth]{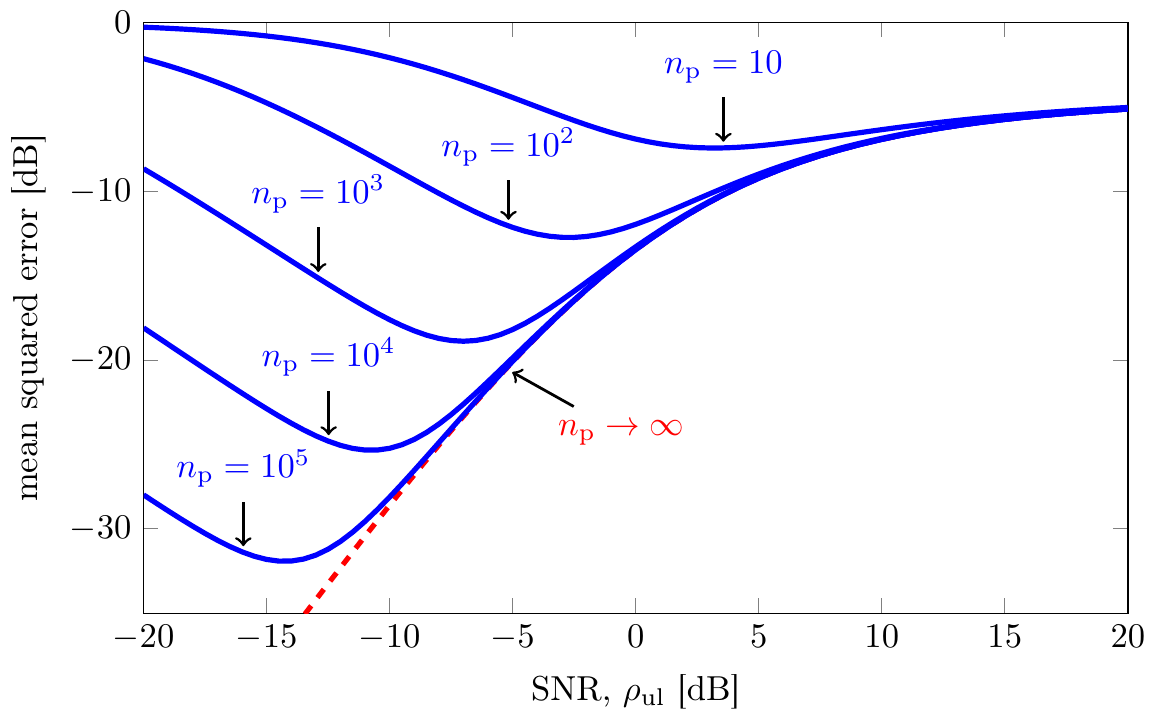}
  \caption{MSE vs.\ SNR for the Bussgang MMSE channel estimator with different pilot-sequence lengths; $1$-bit quantization, $B=100$, $U=1$, and i.i.d. Rayleigh-fading channel.}
  \label{fig:ChEstSNR}
\end{figure}

In Fig.~\ref{fig:ChEstSNR}, we plot the MSE in~\eqref{eq:mmse_1bit} as a function of $\rho_\text{ul}$ for different values of $n\sub{p}$. 
As shown in the figure, for every choice of~$n\sub{p}$, there exists an optimal value of~$\snr\sub{ul}$ that minimizes the MSE.
Reducing $\snr\sub{ul}$ improves performance if $n\sub{p}$ is chosen suitably. 
However, the number of pilot symbols required to benefit from such a reduction increases rapidly as  $\snr\sub{ul}$ is decreased, which makes such an approach impractical. 
Motivated by these observations, we next analyze  the rates achievable when the CSI available at the BS is imperfect. 

\section{Achievable Rates: Imperfect CSI}\label{sec:ar-imperfect-csi}
Let us assume that the BS has an estimate $\widehat\matH$ of $\matH$, which is not necessarily perfect.
Adapting the downlink SINDR expression~\eqref{eq:sindr_downlink} to the case of imperfect CSI at the BS is trivial.
Indeed, it is sufficient to replace $\vecp_u$ in~\eqref{eq:sindr_downlink} by the $u$th column $\widehat\vecp_u$ of $\widehat\matP$
where $\widehat\matP$ denotes the linear precoder used at the BS for the case of imperfect CSI.

Adapting the uplink outage formula~\eqref{eq:ULOUt} is, however, more involved. 
The problem is that in the outage scenario, the received signal (after linear combining) corresponding to the $u$th UE  contains an additional term that depends on the channel-estimation error and is correlated with the useful signal. 
This prevents us from treating this term as additional noise because the result in~\cite[App.~B]{lapidoth02-05a} can no longer be used.

An achievable rate, given $\matH$ and~$\widehat\matH$, can be obtained using the mismatched-decoding framework~\cite{lapidoth02-05a}. 
Specifically, let $\widehat{\matW}$ be~the combining matrix used at the BS for the case of imperfect CSI.
We assume that the BS seeks the codeword transmitted by the $u$th UE that, after being scaled by the ``mismatched'' channel gain $\widehat\vecw^H\matG^{\text{ul}}\widehat{\vech}_u$, is closest to the combiner output corresponding to user $u$th.
A lower bound on the rates achievable by this decoder (under the assumption of Gaussian input and for a given~$\matH$ and $\widehat{\matH}$) can be stated using the generalized mutual information $I^{\text{gmi}}_u$, which, for this scenario, can be computed as follows:
\begin{IEEEeqnarray}{rCl}\label{eq:gmi}
    I^{\text{gmi}}_u &=& -s\bigl(\abs{g-\hat{g}}^2\rho_\text{ul} +\sigma^2\bigr)  \nonumber \\
    &&  +  s \frac{\abs{g}^2\rho_\text{ul}+\sigma^2}{1+s\abs{\hat{g}}^2\rho_\text{ul}} +\ln(1+s\abs{\hat{g}}^2\rho_\text{ul}).  \IEEEeqnarraynumspace
\end{IEEEeqnarray}
Here, 
\begin{IEEEeqnarray}{rCl}
    g &=& \widehat\vecw_u^\text{H} \mathbf{GA} \mathbf{h}_u\\
    \hat{g} &=& \widehat\vecw_u^\text{H} \mathbf{GA} \widehat\vech_u\\
    \sigma^2 &=& \rho_\text{ul}\sum_{v \neq u} \left| \widehat\vecw_u^\text{H} \mathbf{GA} \mathbf{h}_{v}\right|^2+\norm{\mathbf{A}\mathbf{G}^\text{ul}\widehat\vecw_u}^2+\widehat\vecw_u^\text{H}\mathbf{C}_e\widehat\vecw_u \IEEEeqnarraynumspace
\end{IEEEeqnarray}
% \begin{align}
%     &g = \widehat\vecw_u^\text{H} \mathbf{GA} \mathbf{h}_u, \quad
%     \hat{g} = \widehat\vecw_u^\text{H} \mathbf{GA} \widehat\vech_u   \\
%   & \sigma^2 = \rho_\text{ul}\sum_{v \neq u} \left| \widehat\vecw_u^\text{H} \mathbf{GA} \mathbf{h}_{v}\right|^2+\norm{\mathbf{A}\mathbf{G}^\text{ul}\widehat\vecw_u}^2+\widehat\vecw_u^\text{H}\mathbf{C}_e\widehat\vecw_u      
% \end{align}
%
and
\begin{equation}
    s = \frac{-2c+b+\sqrt{b^2+4ac}}{2bc}
\end{equation}
where
\begin{IEEEeqnarray}{rCl}
    a&=&|g|^2\rho_\text{ul}+\sigma^2\\
    b&=&|\hat g|^2\rho_\text{ul}\\
    c&=&|g-\hat g|^2\rho_\text{ul}+\sigma^2.
\end{IEEEeqnarray}

%$a=|g|^2\rho_\text{ul}+\sigma^2$, $b=\rho_\text{ul}|\hat g|^2$, and $c=|g-\hat g|^2\rho_\text{ul}+\sigma^2$.

It turns out that for the case of perfect CSI, $I^{\text{gmi}}_u=\log(1+\gamma_u^\text{ul})$, where~$\gamma_u^\text{ul}$ was defined in~\eqref{eq:sindr_uplink}.
This means that the perfect-CSI rate bound derived in Section~\ref{sec:ar_perfect_csi} follows as special case of~\eqref{eq:gmi}.
Indeed, in the perfect-CSI case, we have that $\hat{g}=g$, which implies that $s=1/\sigma^2$, from which the desired result follows.

% Similar to uplink and following the derivations in \cite{jacobsson2018all}, the outage probability of the $u$th UE with nearest-neighbot decoding, is given by
% \begin{equation} \label{eq:DLOUt}
%     \text{Pr} \{ \operatorname{log}(1+\gamma_u^\text{dl}) < {R} \} 
% \end{equation}
% where $R$ is the target rate. The downlink SINDR $\gamma_u^\text{dl}$ is given by
% \begin{IEEEeqnarray}{rCl} \label{eq:sindr_downlink}
% \gamma_u^\text{dl}  &=& \frac{ \rho_\text{dl} \abs{ \mathbf{h}_u^T  \mathbf{G}^\text{dl} \mathbf{p}_u}^2}{\rho_\text{dl}\sum\limits_{v \neq u} \abs{\mathbf{h}_u^T \mathbf{G}^\text{dl}\mathbf{p}_v}^2 + \mathbf{h}_u^T\mathbf{C}^\text{dl}_{\mathbf{e}}\mathbf{h}_u^* +1}. \IEEEeqnarraynumspace
% \end{IEEEeqnarray}
% where $\mathbf{p}_u$ denotes the $u$th column of the precoding matrix, $\mathbf{G}^\text{dl}$ and $\mathbf{C}_e^\text{dl}$ are the Bussgang matrix and covariance matrix of the quantization noise $\mathbf{e}^\text{dl}$ in downlink. It is worth mentioning that in this formula it is assumed that the UE can estimate the channel gain $\abs{ \mathbf{h}_u^T  \mathbf{G}^\text{dl} \mathbf{p}_u}^2$ perfectly, which can be done at zero cost in the outage scenario by transmitting downlink pilots.
\section{Simulation Results } \label{simulation}
We consider a massive MIMO bidirectional link that operates at a carrier frequency of $30\GHz$. 
We model the antenna array at the BS as a ULA with fixed length $L_\text{ULA}=128\lambda_c=1.28\m$ where $\lambda_c=0.01\m$ is the carrier wavelength. 
We assume that $U=8$ UEs are active and that they are uniformly distributed over a disc with inner radius of $d_\text{min}=50\m$, outer radius of $d_\text{max}=150\m$, and azimuth angle $\phi_u \in (30^{\circ} ,150^{\circ} )$.
Textbook LOS propagation conditions with free-space path loss are considered. 
Furthermore, no user separation is enforced, which makes the results presented in this section rather conservative. 

Throughout this section, we assume a fronthaul rate constraint of $R\sub{fh}=512$\,bit/s/Hz and vary the resolutions of the data converters from $1$ to $8$ bits. 
The corresponding number of BS antennas is obtained as $B = \floor{ R\sub{fh}/(2Q) }$.
Furthermore, the antenna separation is set equal to $L_\text{ULA}/(B\lambda_c)$.
The step size~$\Delta$ of the quantizers is chosen such that the probability that the input signal to the quantizer is clipped is $10^{-4}$. 
For simplicity, we consider MR combining in the uplink and MR precoding in the downlink. 
In the results for the imperfect CSI case, CSI is obtained by means of the Bussgang MMSE estimator with $n\sub{p}=100$ pilot symbols.

As performance metric, we consider the maximum rate at $10\%$ outage probability, which we analyze for the uplink case, for the downlink case, and for the bidirectional link case.
In the last scenario, the maximum rate is defined as the largest rate for which the probability that either the uplink or the downlink are in outage is less than $10\%$.
The SNR values reported in this section refer to a user positioned at the average distance $d_\text{avg}=\frac{2}{3}(d_\text{max}^3-d_\text{min}^3)/(d_\text{max}^2-d_\text{min}^2)$.

\begin{figure}[tp]
\centering
\subfloat[Uplink]{
\includegraphics[width=.78\columnwidth]{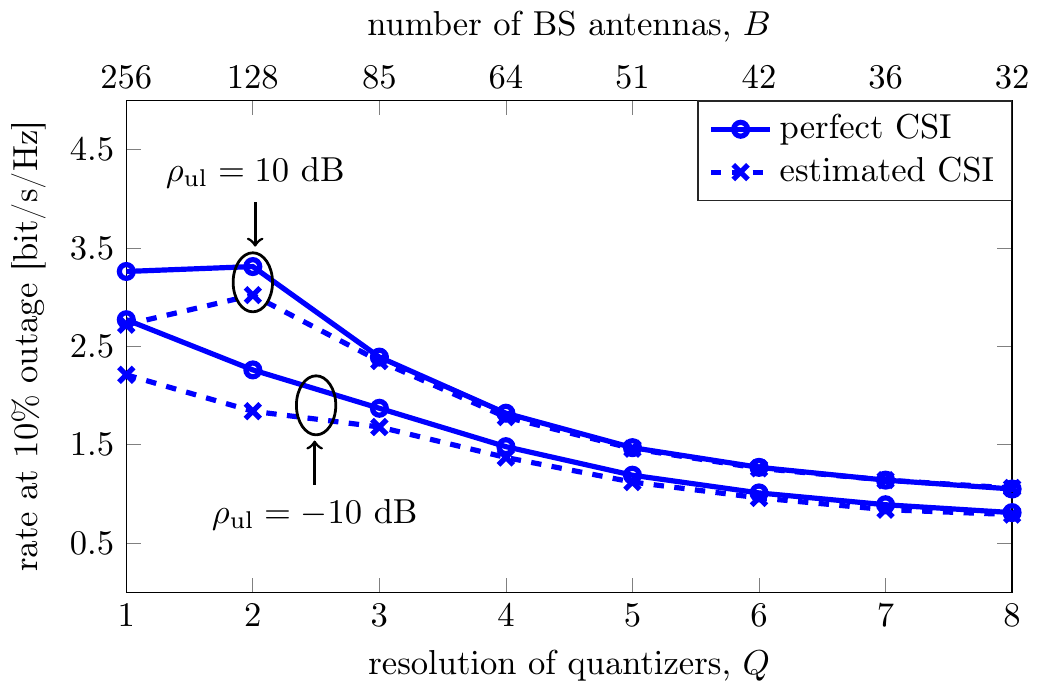}\label{fig:ULRate}}\\
\subfloat[Downlink]{
\includegraphics[width=.78\columnwidth]{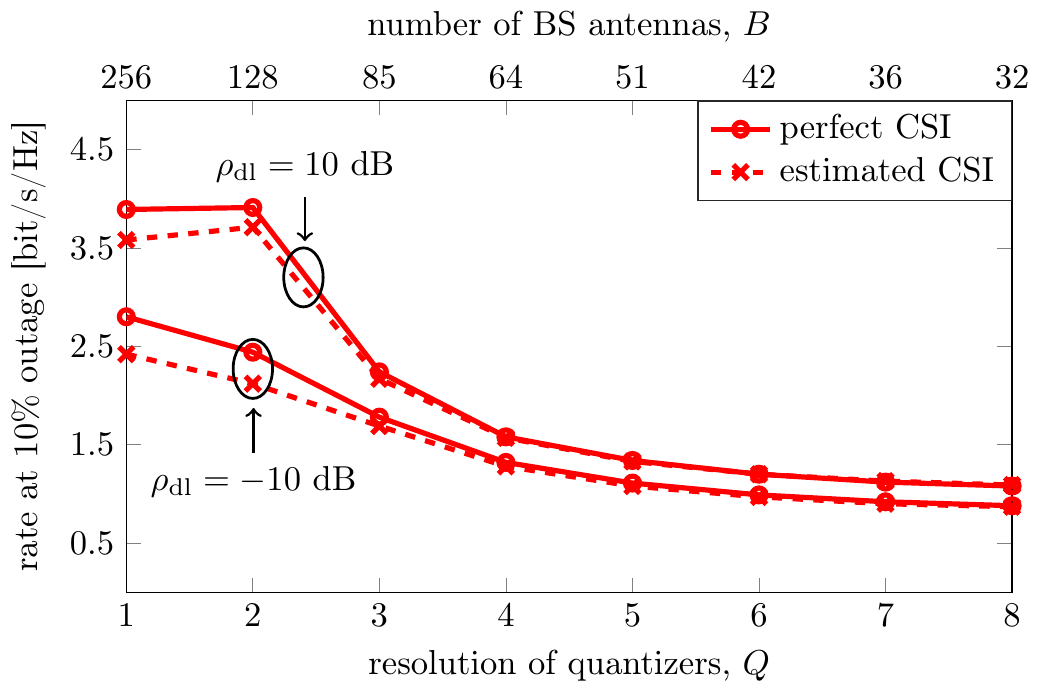}\label{fig:DLRate}}
\caption{Uplink and downlink rates at $10 \%$ outage probability with MR precoding and combining.}
\label{fig:ULDLRate}
\end{figure}

In Fig.~\ref{fig:ULDLRate}, we report the uplink and downlink rates for two SNR values: $-10$\,dB and $10$\,dB.
We see that at low SNR values, the large array gain resulting from the use of $256$ antennas overcomes the performance loss due to the use of $1$-bit quantizers. 
Interestingly, this holds also when the performance loss due to imperfect channel estimation is accounted for.
As the SNR increases, one has to sacrifice half the available antennas for slightly higher (2-bit) resolution.
It is also worth nothing that the performance gap between the perfect and the imperfect CSI cases vanishes rapidly as the resolution of the quantizer increases. 

\begin{figure}[tp]
\centering
\subfloat[Perfect CSI]{
\includegraphics[width=.78\columnwidth]{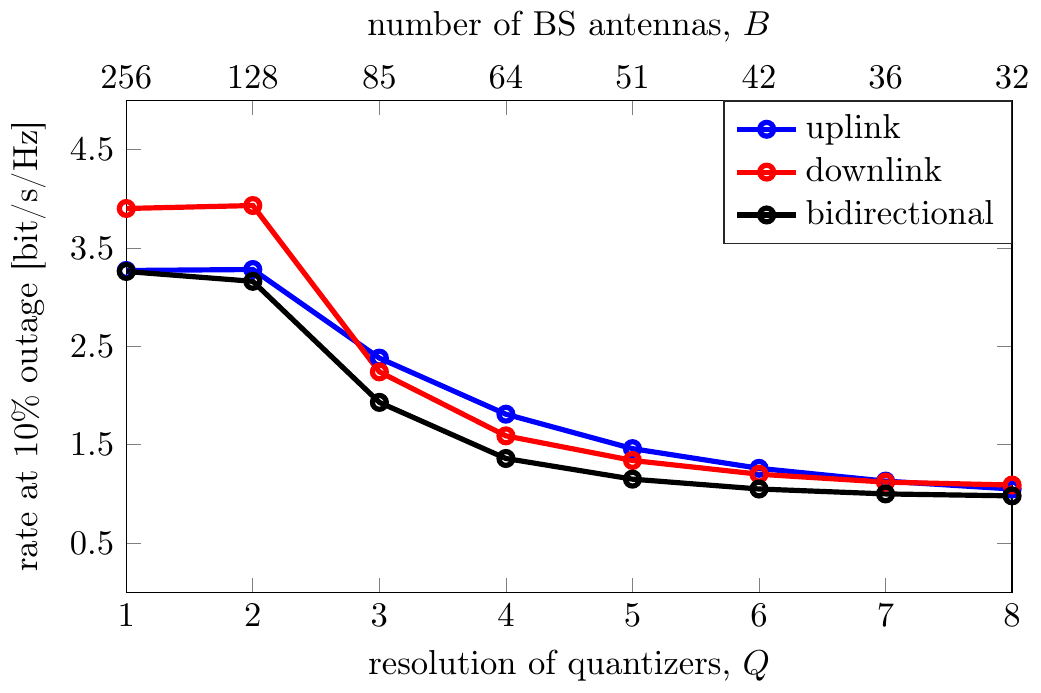}\label{fig:perfectRate}}\\
\subfloat[Imperfect CSI]{
\includegraphics[width=.78\columnwidth]{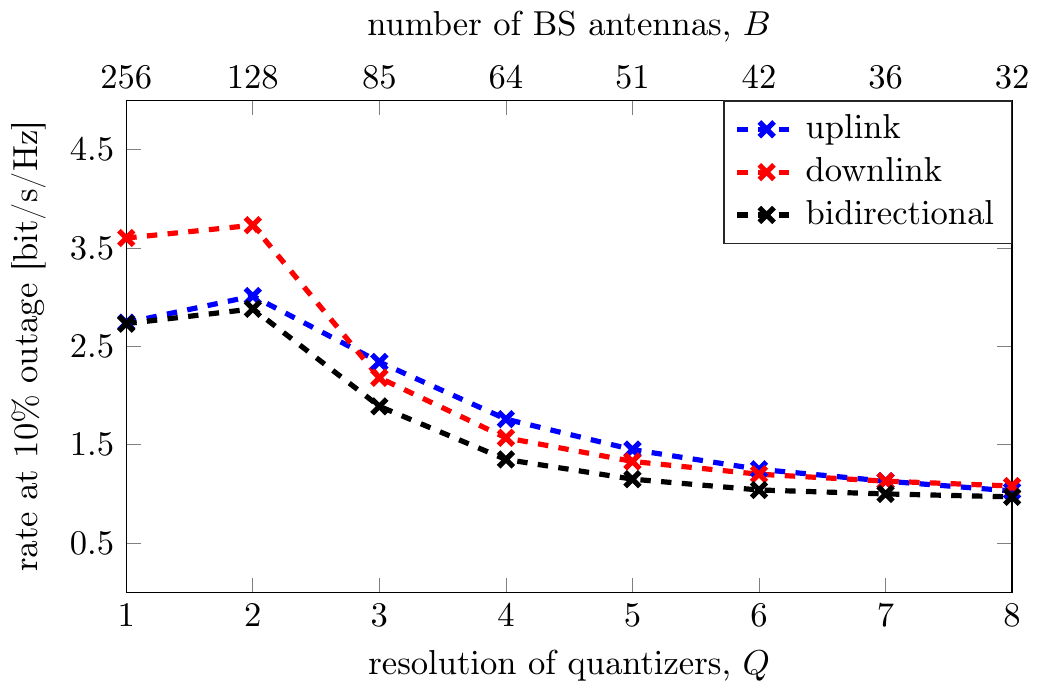}\label{fig:estimatedRate}}
\caption{Uplink, downlink and bi-directional link rates at $10 \%$ outage probability with MR precoding and combining. Both perfect CSI and estimated CSI are considered; $\rho_\text{ul}= 5 \dB$  and $\rho_\text{dl}= 15 
\dB$.}
\label{fig:PerfectImperfectRate}
\end{figure}

In Fig.~\ref{fig:PerfectImperfectRate}, we consider a bidirectional scenario where the uplink operates at $5\dB$ and the downlink operates at $15\dB$. 
As expected, the uplink constitutes the bottleneck in this scenario. Also in this case, the design involving $128$ BS antennas and $2$-bit quantizers results in the highest rate.

% \todo{definition of SNR}
 
% \todo{revised up to here}.

% $\rho_\text{ul}=\rho_\text{dl}=-10$ dB and $\rho_\text{ul}=\rho_\text{dl}=+10$ dB per UE stream at the average distance $d_\text{avg}=\frac{2}{3}(d_\text{max}^3-d_\text{min}^3)/(d_\text{max}^2-d_\text{min}^2)$. We use BLMMSE channel estimator with $n_p=100$ pilot symbols. With enough number of pilots, we can observe that the largest gap between the rate in perfect CSI case and imperfect CSI case occurs at $Q=1$ and further vanishes as $Q$ increases. This results holds in uplink as well as downlink and in high and low SNR.\\
% Finally, we pick a bi-directional scenario and consider bi-directional outage rate as the performance metric to determine the optimal $B$ and $Q$. In Fig \ref{fig:PerfectImperfectRate}, the bi-directional $10 \%$ rate in perfect and imperfect CSI case, as well as uplink and downlink rates are shown. The uplink SNR $\rho_\text{ul}=-5$dB and the downlink SNR $\rho_\text{dl}=-5$dB and the number of pilot symbols is $100$. It can be observed that the optimal point for perfect CSI is at $Q=1$, but it shifts to $Q=2$ in imperfect CSI case which is expected due to limited accuracy in 1bit channel estimation irrespective of $n_p$.

% \begin{figure}[tp]
% \centering
% \subfloat[Perfect CSI]{
% \includegraphics[scale=0.7]{figs/Bidir-perfect.pdf}\label{fig:perfectRate}}\\
% \subfloat[Imperfect CSI]{
% \includegraphics[scale=0.7]{figs/Bidir-imperfect.pdf}\label{fig:estimatedRate}}
% \caption{Uplink, downlink and bi-directional link rates at $10 \%$ outage probability with MR precoding and combining. Both perfect CSI and estimated CSI are considered; $\rho_\text{ul}= 5 \dB$  and $\rho_\text{dl}= 15 
% \dB$.}
% \label{fig:PerfectImperfectRate}
% \end{figure}

\section{Conclusions} \label{conclusion}
We have investigated the design of a fronthaul-constrained full-digital low-precision massive MIMO system.
Our results suggest that architectures involving many antennas with low-precision ($1$--$2$ bit) data converters offer better performance than architectures involving fewer antennas with higher-precision converters.
Extension to our analysis to more realistic propagation conditions and more sophisticated linear precoders, such as the ones in~\cite{bjornson19-02b}, will be presented in future work.
%\balance
\bibliographystyle{IEEEtran}
\bibliography{extracted}

\end{document}